\documentclass[letterpaper, 10 pt, conference]{ieeeconf}  

\IEEEoverridecommandlockouts                              

\usepackage{color}
\usepackage{soul}
\usepackage{graphicx}
\usepackage{subfig}
\usepackage{orcidlink}
\usepackage{booktabs}
\usepackage{multirow}
\usepackage{algorithm}
\usepackage{algpseudocode}

\usepackage[left=0.75in,
            right=0.75in,
            top=0.75in,
            bottom=0.8in]{geometry}

\usepackage{blindtext}

\title{\LARGE \bf
Benchmarking 5G MEC and Cloud infrastructures for planning IoT messaging of CCAM data
}

\author{Felipe Mogollón, Zaloa Fernández\orcidlink{0000-0002-2201-4732}, Josu Pérez and Ángel Martín\orcidlink{0000-0002-1213-6787} \\Vicomtech Foundation, Basque Research and Technology Alliance (BRTA)\\
Paseo Mikeletegi 57, 20009, San Sebasti\'an (Spain) \\ \{fmogollon, zfernandez, jperez, amartin\}@vicomtech.org}

\begin{document}

\maketitle

\begin{abstract}
Vehicles embed lots of sensors supporting driving and safety. Combined with connectivity, they bring new possibilities for Connected, Cooperative and Automated Mobility (CCAM) services that exploit local and global data for a wide understanding beyond the myopic view of local sensors. Internet of Things (IoT) messaging solutions are ideal for vehicular data as they ship core features like the separation of geographic areas, the fusion of different producers on data/sensor types, and concurrent subscription support. Multi-access Edge Computing (MEC) and Cloud infrastructures are key to hosting a virtualized and distributed IoT platform. Currently, the are no benchmarks for assessing the appropriate size of an IoT platform for multiple vehicular data types such as text, image, binary point clouds and video-formatted samples. This paper formulates and executes the tests to get a benchmarking of the performance of a MEC and Cloud platform according to actors' concurrency, data volumes and business levels parameters.
\end{abstract}

\section{INTRODUCTION}

Advanced Driver-Assistance Systems (ADAS) steer the current competition on vehicle manufacturers to enhance the driver experience and safety. This race is bringing lots of sensors to the vehicle's body with autonomous driving on the horizon. These sensors help onboard systems make decision and anticipate environmental situations. However, the next natural leap, as the vehicles get connectivity, is extending onboard sensing with sensors from other vehicles to overcome the physical limitations of onboard sensors. Once local discovery and extension are addressed, the next evolution is based on the ballet and coordination of maneuvers and traffic in an area perspective. Finally, the data from sensors can be exploited in a non-real-time manner for telemetry, diagnostics/maintenance, post-sell, marketing, emergencies and insurance purposes by stakeholders not present on the road and providing Connected, Cooperative and Automated Mobility (CCAM) services.

It becomes evident that all these applications need a large volume of data to be transmitted from multiple devices to different systems interested in the same data flows with a common or different purpose. Internet of Things (IoT) messaging platforms are widely employed solutions for this purpose, bringing maturity, concurrency, scalability, flexibility, specialized data type channels and geographic structure.

5G provides a perfect asset through Multi-access Edge Computing (MEC) to host an IoT messaging platform on top of a virtualized infrastructure \cite{9324778}. This way, data producers and consumers take benefit of the privacy and low-latency processing as close to the sensors as possible. Then a Cloud infrastructure is usually employed to host a central IoT messaging platform managing the hierarchical architecture by aggregating messages from common IoT topics and splitting messages according to the geographical sensors' distribution. In the edge, IoT solutions such as Message Queuing Telemetry Transport (MQTT) or Advanced Message Queuing Protocol (AMQP) are used as they are lightweight for sensors and devices. Then in the Cloud, KAFKA makes the difference with a high-performance, scalable solution that links MQTT and AMQP to act as a global concentrator of local IoT messaging instances.

To fully exploit the scalability potential of the virtualized infrastructures provided in the MEC and the Cloud for such a hierarchical IoT structure, the assessment of resources to cope with the message flows is an essential activity to avoid bottlenecks that will damage the concurrency, capacity or latency service levels.

In the 5G Infrastructure Public Private Partnership (5GPPP) project 5GMETA \cite{seron2022} has been built an IoT platform for vehicular data including, Cooperative Intelligent Transport Systems (C-ITS) JSON and binary formatted messages, individual image recordings and live videos. All of them are embodied in IoT messages allowing live data transmission. It becomes evident that each data format involves a different processing workload and then the required resources will vary. So, 5GMETA implementation applies some cost-effective rules to accommodate different business models.

This paper provides a methodology and framework to benchmark the nominal capacity of an IoT message platform virtualized and operated on top of MEC and Cloud infrastructures. To this end, Section \ref{sota} overviews the IoT architectures for vehicular communications and CCAM applications and the studies on their scalability. Then, Section \ref{setup} explains the 5GMETA platform implementation aspects to consider to maximize processing capacity and minimize latency while fostering a car data marketplace and how the MEC and Cloud setups sustain the developed platform. After, Section \ref{eval} explains the methodology and frameworks employed to score performance, capacity and resulting latency of the IoT platform and the obtained results. Finally, Section \ref{conclusions} highlight the salient results and outlines some future work.

\section{RELATED WORK}
\label{sota}

Industrial processes catalyzed the IoT technologies to deliver sensored data to Supervisory Control And Data Acquisition (SCADA) systems under wired, low throughput and low energy requirements. Fueled by Big Data solutions, the small-scale IoT systems turned into a more dense and connected solution with real-time messaging, data variety, distributed locations and different actors ingredients present in its adoption \cite{9117330}.

These IoT messaging technologies are not only used for communicating the IoT sensors and the process controllers of a pipeline. They are present in other architecture levels, used by management systems for monitoring underlying cellular networks \cite{8767203, 7369627} and virtualization infrastructures \cite{7575883, 9002241}. They employ IoT message solutions to provide data to decision-making systems. Thus, these systems, which goal the robustness of resource-constrained infrastructures, get data with minimal traffic overheads to decide on orchestrating or balancing the allocated assets.

As the mobility, spacial or environmental conditions come into play, wired connectivity of IoT systems is replaced by wireless communications where technologies such as Zigbee, Bluetooth or Long Range Wide Area Network (LoRaWAN), representing Low Power Wide Area Network (LPWAN), work properly for low throughput and energy consumption saving. Then, for higher throughput, WiFi is the general option. However, in scenarios with high mobility and security/privacy requirements, like the CCAM applications, cellular communications make the difference in delivering IoT messages \cite{7380573}. 

Once a cellular network is needed, the suitability of a MEC infrastructure to host IoT message platforms and apply efficiency, granularity, business and privacy rules becomes a natural choice. Additionally, the Radio Network Information Service (RNIS) and the Location Service (LS) are valuable features offered by MEC, as it provides
actual Radio Access Network (RAN) statistics to assess the Quality of Service (QoS) and accurate geo-locations with high-frequency updates \cite{8515152}. 

The application of IoT messaging solutions is so wide that heterogeneous vertical industries are using it to connect data producers and consumers in a versatile and efficient manner. Computation-intensive and delay-sensitive applications like CCAM services require heavy computation tasks processing IoT messages. However, most of the works published in the MEC, IoT and vehicular data handle heavy computing tasks with unlimited processing on local edge or Cloud infrastructures, applying balancing as an efficiency scheme for effectiveness \cite{9238870}. 

The benefits of IoT messaging technology are validated in \cite{9869177} for safety CCAM services with adequate management of regions of interest and geographic data structures. Regarding real-time performance for advanced CCAM applications, authors from \cite{9214633} test the liability of an IoT platform for geo-position-based applications. Regarding the confluence of different stakeholders in the same IoT infrastructure, such as car owners, transit authorities, automobile manufacturers, and other service providers, \cite{9763216} studies the application of blockchain technologies to enforce security. At the same time, \cite{9295328} explores the use of IoT messaging platforms for data monetization built on top of a car data marketplace. Others \cite{9832635} employ blockchain technology hosted at the MEC to agree on Smart Contracts between CCAM applications and cellular network operators for the QoS bridged by the network.

When studying the allocation and consumption of resources in a virtualized infrastructure to cope with the processing of the incoming IoT messages volume, different theoretical models try to reduce the computation workload \cite{8928069, 9826353} and the energy consumption \cite{8647789} of distributed computing infrastructures performing simulations. Some \cite{9819827, 9970003} focus on the connectivity impact to minimize latencies. Others estimate the cost of the solution and applicable business models of the onboard system, excluding the MEC or Cloud platform \cite{9398003}. In terms of virtualization, the evaluation and validation with benchmarks, exploiting Infrastructure-as-a-Service (IaaS) platforms at MEC and Cloud to host an IoT messaging Platform-as-a-Service (PaaS) for vehicular data and CCAM applications is still under-explored.

The evaluation and validation of the cloud acting as a central hub of IoT edge platforms is done in \cite{8325597}. Furthermore, some recent studies also analyze the cost and performance metrics for commercial cloud \cite{9698101} and edge \cite{8605776} infrastructures to handle national-sized IoT platforms. The commercial platforms such as Amazon Web Services (AWS) Greengrass or Azure IoT Edge are evaluated for different data types such as raw audio, raw image and random numbers with user-defined frequency. There are also studies comparing different IoT messaging alternatives \cite{9303425}. The analyzed message queuing systems offer high throughputs and low latencies for processing streamed data. As each one brings a different performance on guaranteed order, reliability, scalability, throughput and latency, the deployment of hierarchical solutions, such as MQTT at the edge and KAFKA at the cloud, or AMQP at the edge and KAFKA at the cloud, are a common good practice in the IoT industry. However, the literature lacks of studies targeting these two-layer architectures. 

The benchmarks for other vertical domains with different parameters and a set of scenarios lead to infrastructure planning and reproducibility for comparison \cite{9921925} and cost reduction \cite{9922500}. The provided toolset in \cite{9826155} emulates a real IoT system mock-up and backs a less biased testing methodology to benchmark IoT applications. At the same time, other authors \cite{8789918} analyze the impact of virtualization on IoT platforms. However, the estimation of necessary computing resources to process different car data types published and subscribed by multiple producers and consumers on top of virtualized IoT solutions hosted in hierarchical MEC and Cloud infrastructures has not yet been formulated or analyzed.

\section{EXPERIMENTATION SETUP}
\label{setup}
\subsection{Hybrid platform}
\begin{figure}[h]
\centering
    \includegraphics[width=0.48\textwidth]{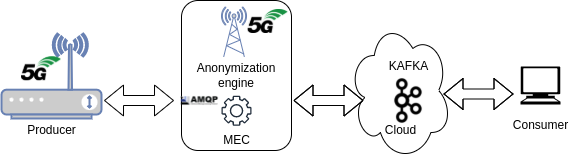}
    \caption{Proposed hybrid platform.}
\label{fig:platform}
\end{figure}

As shown in Fig. \ref{fig:platform}, a hybrid platform has been designed to manage IoT data streams provided by multiple producers and their delivery to the platform's consumers. The main two segments of the platform are a MEC server over a 5G infrastructure that will be the responsible for acquiring IoT data streams from IoT producers. The data service at the MEC anonymizes those IoT data streams in order to be sent to the next section of the proposed platform, a cloud deployment on a commercial provider. The cloud data service is responsible for answering the data requests from consumers to get the appropriate IoT streams from the selected MEC where consumers push data. Hardware characteristics for both segments can be seen on Table \ref{tab:PlatforHardware}

The main software engine in the MEC is an AMQP broker that is connected to a KAFKA topic on the cloud section of the infrastructure. This KAFKA engine in the cloud infrastructure can be commanded by a consumer to select the appropriate topic from KAFKA where anonymized data can be read.


\begin{table}[h]
\centering
\caption{Hybrid platform hardware characteristics.}
\begin{tabular}[t]{lccccc} 
\toprule
\textbf{Segment} & \textbf{Number of CPUs} & \textbf{RAM} \\
\midrule
\textbf{MEC} & 36 & 128 GB  \\ 
\textbf{Cloud} & 4 & 16 GB  \\ 
\bottomrule
    \end{tabular}
    \label{tab:PlatforHardware}
\end{table}

\subsection{Anonymization schemes}

Three anonymization schemes have been defined in the MEC section of the platform. The alternatives allow to control the required computing resources in the MEC and the data throughput provided to consumers. Thus, those schemes are related to the capabilities of hardware resources provided to the anonymization engine and the number of samples anonymized and pipelined. The schemes are compiled in Table \ref{tab:AnonymizationScenarios}.

\begin{table}[h]
\centering
\caption{Anonymization schemes.}
\begin{tabular}[t]{lccccc} 
\toprule
\textbf{Scheme} & \textbf{Number of CPUs} & \textbf{RAM in GB} & \textbf{Sampling rate} \\
\midrule
\textbf{Small} & 2 & 2 & 1 message / 5 sec \\ 
\textbf{Medium} & 4 & 4 & 1 message / sec \\
\textbf{Large}  & 8 & 8 & No sampling   \\ 
\bottomrule
    \end{tabular}
    \label{tab:AnonymizationScenarios}
\end{table}

\subsection{Local low resource platform}
\label{local_platform}
\begin{figure}[h]
\centering
    \includegraphics[width=0.48\textwidth]{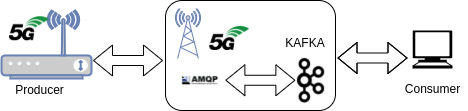}
    \caption{Local resource platform.}
\label{fig:local_platform}
\end{figure}

As shown in Fig. \ref{fig:local_platform}, a local resource platform has been designed in order to not have any dependence on cloud providers or Internet infrastructure.
That local platform has the same logic components, excluding the anonymization engine, as the hybrid platform but runs on a single computer instead of a MEC and cloud combination. The anonymization engine's absence is explained by the lack of computing resources compared with the hybrid platform. In Table \ref{tab:LocalPlatformHardware}, hardware characteristics for this platform can be seen to understand that absence compared to hybrid platform hardware shown in Table \ref{tab:PlatforHardware}.

\begin{table}[h]
\centering
\caption{Local resource platform hardware characteristics.}
\begin{tabular}[t]{lccccc} 
\toprule
\textbf{Segment} & \textbf{Number of CPUs} & \textbf{RAM in GB} \\
\midrule
\textbf{Local resource platform} & 20 & 16  \\ 
\bottomrule
    \end{tabular}
    \label{tab:LocalPlatformHardware}
\end{table}

\subsection{Producer}
A custom producer has been created for both platforms (for the hybrid one and for the local one) to push data to the platform.
This producer will generate customized C-ITS messages with producer IDs and timestamps and will ask the cloud for which MEC to send data. Then the producer will send the C-ITS data stream through an AMQP broker. 

\subsection{Consumer}
As in the case of the producer, for both platforms (the hybrid one and the local one), a custom consumer has been created to receive the data generated by the producer to perform some calculations about the platform's performance.

This consumer will request a KAFKA topic from the cloud to consume data from a given MEC, where consumers will push data. Once the consumer receives the appropriate topic and address of KAFKA service, will connect to that topic and will start consuming data by logging the received Cooperative Intelligent Transport Systems (C-ITS) \cite{cits} messages and annotating them with the current timestamp.

\subsection{5G infrastructure}
For the deployment of both 5G Core (5GC) and 5G New Radio (5GNR) base station (gNB) in Stand Alone (SA) mode, an AMARI Callbox Pro provided by Amarisoft has been used. Amarisoft is hardware agnostic in terms of supported hardware, as it can work with different hardware equipment to provide the necessary radio connectivity. The 5G network has been deployed in Standalone (SA) mode and has reduced coverage.


\section{EVALUATION}
\label{eval}
\subsection{Methodology and metrics}

\begin{figure}[h]
\centering
    \includegraphics[width=0.48\textwidth]{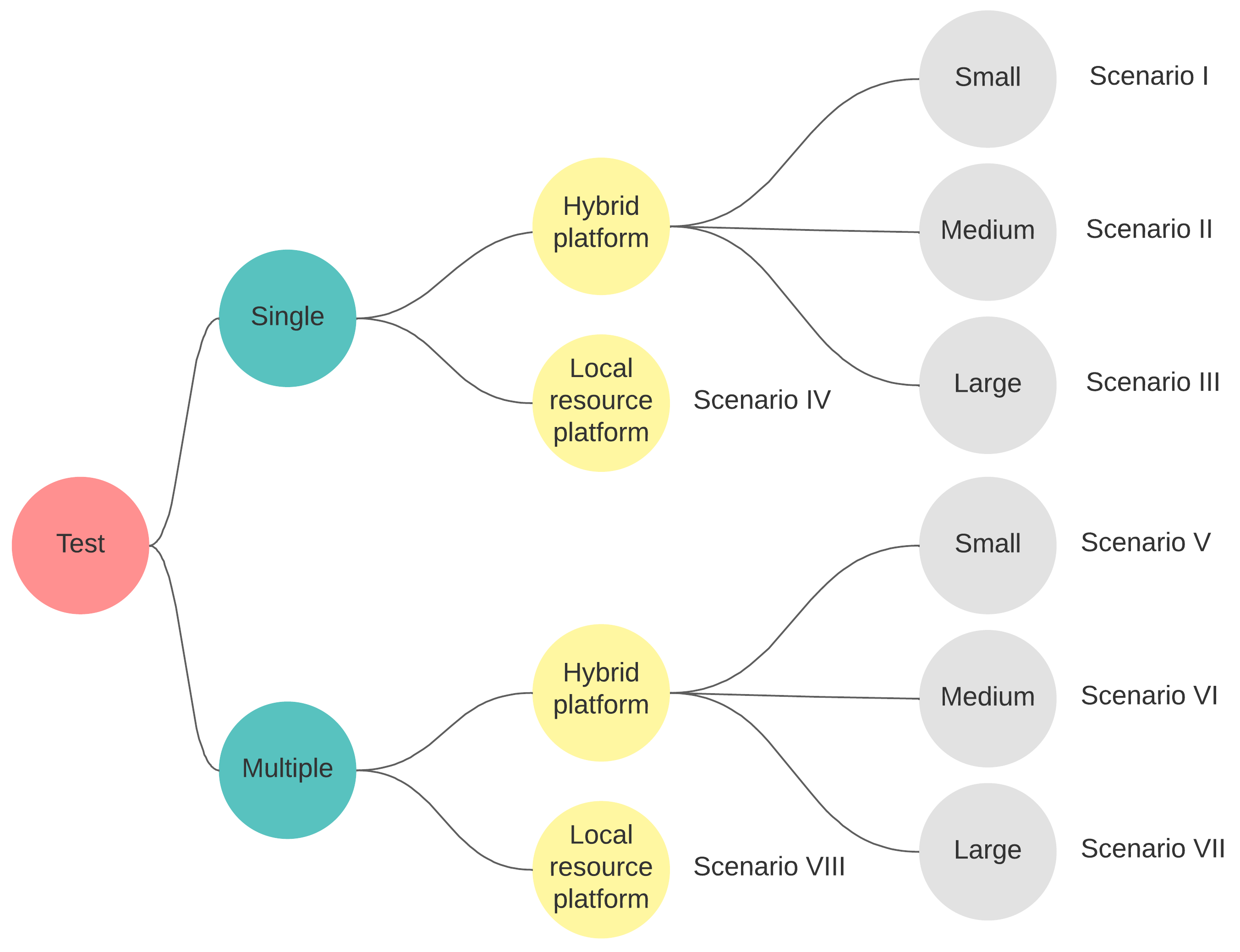}
    \caption{Scenarios defined for platform benchmarking.}
\label{fig:scenarios}
\end{figure}

This section presents the setup and methodology deployed to evaluate the different platforms. Fig. \ref{fig:scenarios} shows the eight scenarios considered for assessing the hybrid and local low resource platforms in combination with the three anonymization schemes defined in Section \ref{setup}.

\label{datainput}
In a single scenario, we have considered one producer for each test; in a multiple scenario, we have considered ten simultaneous producers. Moreover, ten tests of 10 minutes each have been carried out in each scenario. In addition, each producer sends C-ITS messages at a 2 Hz frequency. The transmitted C-ITS has a payload of 1280 bytes.

Considering these eight scenarios, several measurements have been carried out to obtain the application latency and the packet losses. In addition, the performance of the 5G infrastructure that is connected to the platform has also been measured. For this purpose, the Round-trip time (RTT) and the throughput of the 5G infrastructure have been measured. Some custom and proprietary logging tools have been used to obtain these measurements, some embedded and others external to the applications under test. 

To carry out the measurement related to the 5G infrastructure, we have used Dekra's TACS4 Performance Tool. This tool enables simultaneous performance testing and user experience analysis of wired and wireless access networks for voice and data services. The duration of each test has been set at 120 seconds, obtaining one sample for every second, thus reducing the variability in the result. Moreover, each test has been repeated five times. 

Developed producers and consumers have been used to evaluate the performance of both hybrid and local platforms. Latency and packet losses have been obtained following Algorithm \ref{alg:latency} and Algorithm \ref{alg:packetloss} respectively.\\

\begin{algorithm}
\caption{Algorithm for obtaining the latency of C-ITS messages.}\label{alg:latency}
\begin{algorithmic}[1]
\While {NewCITSMessage}
    \State $\mathit{getCurrentTime()}$
    \State $\mathit{getOriginTimefromMessage()}$
  \State $\mathit{latency} = CurrentTime - OriginTime$
\EndWhile
\State \textbf{end}
\end{algorithmic}
\end{algorithm}

\begin{algorithm}
\caption{Algorithm for obtaining the packet losses of C-ITS messages.}\label{alg:packetloss}
\begin{algorithmic}[1]

\State $\mathit{loadReceivedMessages()}$
\State $\mathit{loadSentMessages()}$
  \ForAll{$\mathit{message}\in receivedMessage$}
  \State $\mathit{IdTime = getOriginTimeAndId(message)}$
    \If{$\mathit{IdTime}$ in $\mathit{sentMessages}$}
      \State  $\mathit{messageReceived}$
    \Else
      \State $\mathit{messageLost}$
    \EndIf
  \EndFor
\State \textbf{end}
\end{algorithmic}
\end{algorithm}

Finally, both the producer and consumer clocks are synchronized by Network Time Protocol (NTP).

\subsection{Results}
First, to understand the radio limits of the 5G testing setup, Table \ref{tab:Throughput5G} provides the results of the throughput measurements on the uplink (UL) and downlink (DL) of the deployed 5G infrastructure. Based on the configuration of the network, the achievable DL throughput is close to 150 Mbit/s (below the 190 Mbit/s when there is only DL traffic transmitting) and UL throughput is close to 100 Mbit/s (below the 120 Mbit/s when there is only UL traffic transmitting). As seen in Table \ref{tab:Throughput5G}, the performance in DL is slightly higher. It can be seen that, on average, in DL, a throughput of 152 Mbit/s is achieved, while in UL, a throughput of 100 Mbit/s is achieved. Nevertheless, during the measurements, values of 187 Mbit/s were reached in DL and 113 Mbit/s in UL.

\begin{table}[h]
\centering
\caption{Platform performance in relation to the 5G infrastructure.}
\begin{tabular}[t]{lcc} 
\toprule
\multicolumn{1}{l}{\textbf{Metric}} & \textbf{Throughput (Mbit/s)} \\
\midrule
\textbf{DL average} & 151.88 \\ 
\textbf{DL peak} & 172.95\\ 
\textbf{DL maximum} & 187.25\\ 
\textbf{UL average} & 99.79\\ 
\textbf{UL peak} & 109.66\\ 
\textbf{UL maximum} & 113.58\\ 
\bottomrule
    \end{tabular}
    \label{tab:Throughput5G}
\end{table}

\begin{figure}[h]
\centering
    \includegraphics[width=0.48\textwidth]{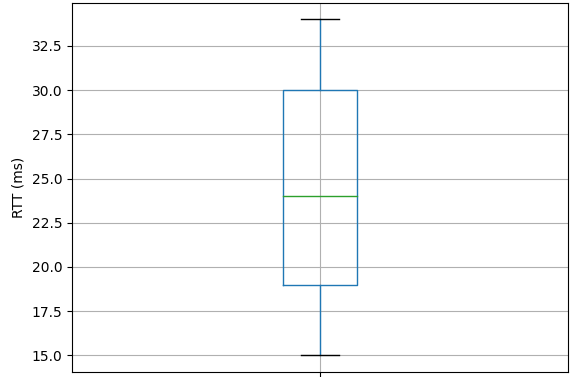}
    \caption{Platform RTT in relation to the 5G infrastructure.}
\label{fig:ping5gresult}
\end{figure}

Fig. \ref{fig:ping5gresult} shows the obtained RTT from 5G infrastructure to the proposed platform. As can be seen, the average value of the RTT is about 25 ms, but measured values range from 15 ms to 34 ms. Considering this is a non-commercial 5G network and an experimental platform, the RTT values obtained are not very high and remain stable within a small range.

\begin{figure}[h]
\centering
    \includegraphics[width=0.48\textwidth]{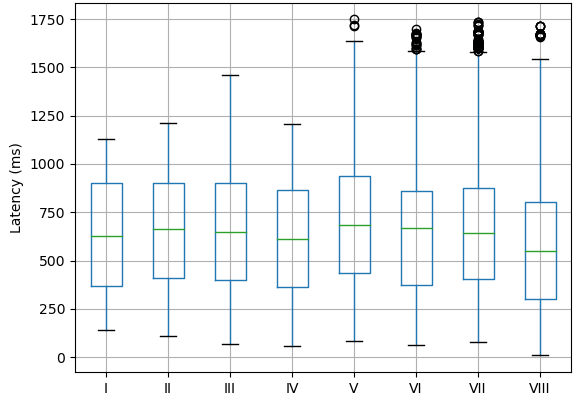}
    \caption{Application latency of platforms for all scenarios.}
\label{fig:lateny_summary}
\end{figure}

As shown in Fig. \ref{fig:lateny_summary} and Table \ref{tab:LatencyAndPER}, all latency parameters like mean, median and deviation ($\sigma$) are quite similar in all the evaluated scenarios. It can be noticed that the mean latency in scenarios IV and VIII is lower than in other scenarios. These scenarios are related to the local low resources platform, which, as mentioned in Section \ref{local_platform}, does not have any anonymization engine because of the lack of computing resources. That leads the platform to a direct transmission between the AMQP topic, which collects C-ITS messages from producers and the KAFKA topic, where consumers get those data streams. This means that the anonymization process that happens only in a hybrid platform introduces some delay but a very small one, as it can be seen comparing results between scenarios with anonymization (I, II, III, V, VI, VII) and the scenarios without it (IV and VIII).\\

\begin{table}[h]
\centering
\caption{Benchmarking of platforms for all scenarios.}
\begin{tabular}[t]{lccccc} 
\toprule
\multicolumn{1}{c}{\multirow{2}{*}{\textbf{Scenario}}} & \multicolumn{3}{c}{\textbf{Latency}} & \multicolumn{1}{c}{\multirow{2}{*}{\textbf{Packet loss (\%)}}} \\
\multicolumn{1}{c}{} & \multicolumn{1}{c}{\textbf{Mean (ms)}} & \multicolumn{1}{c}{\textbf{Median (ms)}} & \multicolumn{1}{c}{\textbf{{$\sigma$} (ms)}} & \multicolumn{1}{c}{} \\
\midrule
(\textbf{I}) & 637.64 & 626.5 & 289.17 & 90.01\\ 
(\textbf{II}) & 653.82 & 661 & 287.5  & 50.25\\
(\textbf{III})  & 648.07 & 647 & 290.48 & 0.68  \\ 
(\textbf{IV}) & 609.11 & 609 & 288.74  & 0.52 \\ 
(\textbf{V}) & 682.86 & 684.5 & 291.15  & 91.29 \\ 
(\textbf{VI}) & 627.33 & 673 & 285.47   & 56.64\\ 
(\textbf{VII}) & 648.83 & 650 & 277.09  & 13.27 \\ 
(\textbf{VIII}) & 553.86 & 552 & 291.6  & 28.81 \\ 
\bottomrule
    \end{tabular}
    \label{tab:LatencyAndPER}
\end{table}

On the other side, we see the most significant difference between the evaluated scenarios in the measure of packet loss. Specifically, the scenarios with anonymization schemes show noticeable differences in the packet loss metrics, as shown in Table \ref{tab:LatencyAndPER}. This can be explained by the sampling parameter shown in Table \ref{tab:AnonymizationScenarios}. There we can see that the Small and Medium scenarios have some subsampling of the input stream. This will lead to packet loss. \\
As said in Section \ref{datainput}, input data has a frequency of 2 Hz, which leads to, taking into account the sampling parameters defined in Table \ref{tab:AnonymizationScenarios}, a predicted loss shown in Table \ref{tab:predictedlosses}. In order to obtain those predicted packet loss values, equation \ref{eq:predictor} is used, where $P_{l}$ is the predicted packet loss, $S_{r}$ is the sampling rate and $F_{data}$ is the frequency of input data.

\begin{equation}
    P_{l} = \left (1 - \frac{S_{r}}{F_{data}} \right ) \cdot 100
    \label{eq:predictor}
\end{equation}



\begin{table}[h]
\centering
\caption{Predicted packet losses}
\begin{tabular}[t]{lcc} 
\toprule
\multicolumn{1}{l}{\textbf{Scenario}} & \textbf{Predicted packet losses} \\
\midrule
\textbf{Small} & 90 \% \\ 
\textbf{Medium} & 50 \% \\ 
\textbf{Large} & 0 \% \\ 
\bottomrule
    \end{tabular}
    \label{tab:predictedlosses}
\end{table}

Comparing the packet losses from Tables \ref{tab:LatencyAndPER} and \ref{tab:predictedlosses}, single scenarios (I, II, III) are quite similar, and in multiple scenarios (V, VI, VII), there is a light overhead. That overhead can be explained because, in those scenarios, multiple instances for producers were used, and the platform may have some saturation. This leads the consumers to cannot get all the stream data.
Regarding this, in scenarios IV and VIII, this saturation when a single producer or multiple producers are pushing data is more significant because, in those scenarios, the platform used is the one with low resources. \\
From the point of view of a sensitivity analysis, it can be seen that the sampling rate parameter from Table \ref{tab:AnonymizationScenarios} has a direct impact on both the predicted packet loss from Table \ref{tab:predictedlosses} and the obtained packet loss during the benchmarking of the platform from Table \ref{tab:LatencyAndPER} as shown in equation \ref{eq:predictor}.

\section{CONCLUSIONS AND FUTURE WORK}
\label{conclusions}

Vehicles are embracing lots of technologies, from sensing to ADAS and infotainment systems. With the universal integration of cellular connectivity capacity, the vehicles will be ready to join a cooperative community for traffic efficiency and safety. And this is only the beginning. The upcoming innovations on the horizon, i.e., the next generation of CCAM applications, will find new ways of monetizing car data from different and new business models. To sustain this new industry on top of vehicle data, IoT messaging is key to connecting data producers and consumers using sophisticated edge and cloud infrastructures to have virtually unlimited resources to process vast volumes of data. The scalability potential of MEC and Cloud infrastructures is not for free, and it is important to be efficient and appropriately estimate the size of the IoT messaging platform to deliver all the data with their different formats and rates.

This paper provides results on stressing a hierarchical MEC and Cloud IoT messaging platform connecting data from producers to consumers. The results show that the application latency remains stable in all the scenarios analyzed and that the anonymization schemes add a reduced latency to the platform. As for packet loss, we conclude that the obtained packet losses are as expected due to the anonymization schemes, but there is a bottleneck when working with multiple scenarios.

In future work, on the one hand, we have to analyze the capacity of the multiple scenarios to optimize their performance. On the other hand, we have to evaluate the proposed platform in different conditions and types of messages.




\addtolength{\textheight}{0cm}   




\section*{ACKNOWLEDGMENT}

This research was supported by the European Union’s Horizon 2020 research and innovation program under grant agreement No. 957360 (5GMETA project). Authors thank DEKRA Spain for the TACS4 Performance Test Tool license.


\bibliographystyle{IEEEtran}
\bibliography{main.bib}

\begin{thebibliography}{10}
\providecommand{\url}[1]{#1}
\csname url@samestyle\endcsname
\providecommand{\newblock}{\relax}
\providecommand{\bibinfo}[2]{#2}
\providecommand{\BIBentrySTDinterwordspacing}{\spaceskip=0pt\relax}
\providecommand{\BIBentryALTinterwordstretchfactor}{4}
\providecommand{\BIBentryALTinterwordspacing}{\spaceskip=\fontdimen2\font plus
\BIBentryALTinterwordstretchfactor\fontdimen3\font minus
  \fontdimen4\font\relax}
\providecommand{\BIBforeignlanguage}[2]{{%
\expandafter\ifx\csname l@#1\endcsname\relax
\typeout{** WARNING: IEEEtran.bst: No hyphenation pattern has been}%
\typeout{** loaded for the language `#1'. Using the pattern for}%
\typeout{** the default language instead.}%
\else
\language=\csname l@#1\endcsname
\fi
#2}}
\providecommand{\BIBdecl}{\relax}
\BIBdecl

\bibitem{9324778}
L.~Nkenyereye, J.~Hwang, Q.-V. Pham, and J.~Song, ``Virtual iot service slice
  functions for multiaccess edge computing platform,'' \emph{IEEE Internet of
  Things Journal}, vol.~8, no.~14, pp. 11\,233--11\,248, 2021.

\bibitem{seron2022}
M.~Seron, A.~Martin, and G.~Velez, ``Life cycle management of automotive data
  functions in mec infrastructures,'' in \emph{2022 IEEE Future Networks World
  Forum (FNWF)}.\hskip 1em plus 0.5em minus 0.4em\relax IEEE, 2022, pp.
  407--412.

\bibitem{9117330}
M.~H. Alquwatli, M.~H. Habaebi, and S.~Khan, ``Review of scada systems and iot
  honeypots,'' in \emph{2019 IEEE 6th International Conference on Engineering
  Technologies and Applied Sciences (ICETAS)}, 2019, pp. 1--6.

\bibitem{8767203}
B.~Mostafa, ``Monitoring internet of things networks,'' in \emph{2019 IEEE 5th
  World Forum on Internet of Things (WF-IoT)}, 2019, pp. 295--298.

\bibitem{7369627}
S.~Hu, ``Dynamic monitoring based on wireless sensor networks of iot,'' in
  \emph{2015 International Conference on Logistics, Informatics and Service
  Sciences (LISS)}, 2015, pp. 1--4.

\bibitem{7575883}
A.~Karapantelakis, H.~Liang, K.~Wang, K.~Vandikas, R.~Inam, E.~Fersman,
  I.~Mulas-Viela, N.~Seyvet, and V.~Giannokostas, ``Devops for iot applications
  using cellular networks and cloud,'' in \emph{2016 IEEE 4th International
  Conference on Future Internet of Things and Cloud (FiCloud)}, 2016, pp.
  340--347.

\bibitem{9002241}
R.~Sasvanth~Narayan, V.~Loganathan, P.~Lakkar, and B.~Suganthan, ``Iot cloud
  based optimization of vehicle using monitoring systems,'' in \emph{2019
  International Conference on Communication and Electronics Systems (ICCES)},
  2019, pp. 1108--1112.

\bibitem{7380573}
K.~M., S.~M., and R.~Banakar, ``Evolution of iot in smart vehicles: An
  overview,'' in \emph{2015 International Conference on Green Computing and
  Internet of Things (ICGCIoT)}, 2015, pp. 804--809.

\bibitem{8515152}
F.~Giust, V.~Sciancalepore, D.~Sabella, M.~C. Filippou, S.~Mangiante,
  W.~Featherstone, and D.~Munaretto, ``Multi-access edge computing: The driver
  behind the wheel of 5g-connected cars,'' \emph{IEEE Communications Standards
  Magazine}, vol.~2, no.~3, pp. 66--73, 2018.

\bibitem{9238870}
L.~Yang, H.~Yao, X.~Zhang, J.~Wang, and Y.~Liu, ``Multi-uav deployment for mec
  enhanced iot networks,'' in \emph{2020 IEEE/CIC International Conference on
  Communications in China (ICCC)}, 2020, pp. 436--441.

\bibitem{9869177}
A.~S. Canto, P.~V. Matias, R.~O. Moreira, T.~A. Sampaio, A.~E. Santos, D.~F.
  Luiz, C.~B. Carvalho, and W.~S. Júnior, ``A mobile iot system for the
  detection and prevention of vehicular collisions,'' in \emph{2022 IEEE
  International Conference on Consumer Electronics - Taiwan}, 2022, pp.
  511--512.

\bibitem{9214633}
A.~H. Alquhali, M.~Roslee, M.~Y. Alias, and K.~S. Mohamed, ``Iot based
  real-time vehicle tracking system,'' in \emph{2019 IEEE Conference on
  Sustainable Utilization and Development in Engineering and Technologies
  (CSUDET)}, 2019, pp. 265--270.

\bibitem{9763216}
A.~L. Shrivastava and R.~K. Dwivedi, ``A secure design of the smart vehicular
  iot system using blockchain technology,'' in \emph{2022 9th International
  Conference on Computing for Sustainable Global Development (INDIACom)}, 2022,
  pp. 616--620.

\bibitem{9295328}
T.~Sutjarittham, H.~H. Gharakheili, S.~S. Kanhere, and V.~Sivaraman,
  ``Monetizing parking iot data via demand prediction and optimal space
  sharing,'' \emph{IEEE Internet of Things Journal}, vol.~9, no.~8, pp.
  5629--5644, 2022.

\bibitem{9832635}
A.~Ahmed, S.~Abdullah, S.~Iftikhar, I.~Ahmad, S.~Ajmal, and Q.~Hussain, ``A
  novel blockchain based secured and qos aware iot vehicular network in edge
  cloud computing,'' \emph{IEEE Access}, vol.~10, pp. 77\,707--77\,722, 2022.

\bibitem{8928069}
H.~Xu, C.~Pan, K.~Wang, M.~Chen, and A.~Nallanathan, ``Resource allocation for
  uav-assisted iot networks with energy harvesting and computation
  offloading,'' in \emph{2019 11th International Conference on Wireless
  Communications and Signal Processing (WCSP)}, 2019, pp. 1--7.

\bibitem{9826353}
M.~A. Hossain, A.~R. Hossain, and N.~Ansari, ``Numerology-capable uav-mec for
  future generation massive iot networks,'' \emph{IEEE Internet of Things
  Journal}, vol.~9, no.~23, pp. 23\,860--23\,868, 2022.

\bibitem{8647789}
Y.~Du, K.~Wang, K.~Yang, and G.~Zhang, ``Energy-efficient resource allocation
  in uav based mec system for iot devices,'' in \emph{2018 IEEE Global
  Communications Conference (GLOBECOM)}, 2018, pp. 1--6.

\bibitem{9819827}
G.~Manogaran, J.~Gao, and T.~N. Nguyen, ``Optimizing resource and service
  allocations for iot-assisted intelligent transportation systems,'' \emph{IEEE
  Transactions on Intelligent Transportation Systems}, pp. 1--11, 2022.

\bibitem{9970003}
G.~Yang and Y.~Yao, ``Resource allocation control of uav-assisted iot
  communication device,'' \emph{IEEE Transactions on Intelligent Transportation
  Systems}, pp. 1--9, 2022.

\bibitem{9398003}
M.~S.~A. khan, T.~Akter, M.~M.~Y. Hossain, M.~S. Hossain, S.~Mazumder, and
  M.~K. Alam, ``Design and business modeling of an iot based cost-effective
  vehicular monitoring system for next generation smart vehicle,'' in
  \emph{2020 IEEE International Women in Engineering (WIE) Conference on
  Electrical and Computer Engineering (WIECON-ECE)}, 2020, pp. 440--443.

\bibitem{8325597}
A.~C. Marosi, R.~Lovas, A.~Kisari, and E.~Simonyi, ``A novel iot platform for
  the era of connected cars,'' in \emph{2018 IEEE International Conference on
  Future IoT Technologies (Future IoT)}, 2018, pp. 1--11.

\bibitem{9698101}
L.~Calderoni, D.~Maio, and L.~Tullini, ``Benchmarking cloud providers on
  serverless iot back-end infrastructures,'' \emph{IEEE Internet of Things
  Journal}, vol.~9, no.~16, pp. 15\,255--15\,269, 2022.

\bibitem{8605776}
A.~Das, S.~Patterson, and M.~Wittie, ``Edgebench: Benchmarking edge computing
  platforms,'' in \emph{2018 IEEE/ACM International Conference on Utility and
  Cloud Computing Companion (UCC Companion)}, 2018, pp. 175--180.

\bibitem{9303425}
G.~Fu, Y.~Zhang, and G.~Yu, ``A fair comparison of message queuing systems,''
  \emph{IEEE Access}, vol.~9, pp. 421--432, 2021.

\bibitem{9921925}
H.~Krasowski and M.~Althoff, ``Commonocean: Composable benchmarks for motion
  planning on oceans,'' in \emph{2022 IEEE 25th International Conference on
  Intelligent Transportation Systems (ITSC)}, 2022, pp. 1676--1682.

\bibitem{9922500}
C.~Alias and J.~Z. Felde, ``Evaluating the economic performance of a
  decentralized waterborne container transportation service using autonomous
  inland vessels,'' in \emph{2022 IEEE 25th International Conference on
  Intelligent Transportation Systems (ITSC)}, 2022, pp. 3571--3576.

\bibitem{9826155}
R.~Cristea, M.~Feraru, and C.~Paduraru, ``Building blocks for iot testing - a
  benchmark of iot apps and a functional testing framework,'' in \emph{2022
  IEEE/ACM 4th International Workshop on Software Engineering Research and
  Practices for the IoT (SERP4IoT)}, 2022, pp. 25--32.

\bibitem{8789918}
S.~Venticinque, ``Benchmarking physical and virtual iot platforms,'' in
  \emph{2019 IEEE International Conference on Cloud Engineering (IC2E)}, 2019,
  pp. 247--252.

\bibitem{cits}
\BIBentryALTinterwordspacing
{ETSI Technical Committee Intelligent Transport Systems (ITS)},
  ``\BIBforeignlanguage{en}{Intelligent transport systems (its) vehicular
  communications basic set of applications part 2: Specification of cooperative
  awareness basic service},'' European Telecommunications Standards Institute,
  Standard ETSI EN 302 637-2 V1.3.1 (2014-09), 2014. [Online]. Available:
  \url{https://www.etsi.org/deliver/etsi_en/302600_302699/30263702/01.03.01_30/en_30263702v010301v.pdf}
\BIBentrySTDinterwordspacing

\end{thebibliography}

\end{document}